# BUILDING A DATA WAREHOUSE FOR NATIONAL SOCIAL SECURITY FUND OF THE REPUBLIC OF TUNISIA

## Mohamed Salah GOUIDER[1], Amine FARHAT[2]

BESTMOD Laboratory – Institut Supérieur de Gestion

41, rue de la liberté, cite Bouchoucha

Bardo, 2000, Tunis, TUNISIA

ms.gouider@isg.rnu.tn[1], farhat_amine@yahoo.fr[2]

## *ABSTRACT*

*The amounts of data available to decision makers are increasingly important, given the network availability, low cost storage and diversity of applications. To maximize the potential of these data within the National Social Security Fund (NSSF) in Tunisia, we have built a data warehouse as a multidimensional database, cleaned, homogenized, historicized and consolidated. We used Oracle Warehouse Builder to extract, transform and load the source data into the Data Warehouse, by applying the KDD process. We have implemented the Data Warehouse as an Oracle OLAP. The knowledge extraction has been performed using the Oracle Discoverer tool. This allowed users to take maximum advantage of knowledge as a regular report or as ad hoc queries. We started by implementing the main topic for this public institution, accounting for the movements of insured persons. The great success that has followed the completion of this work has encouraged the NSSF to complete the achievement of other topics of interest within the NSSF. We suggest in the near future to use Multidimensional Data Mining to extract hidden knowledge and that are not predictable by the OLAP.*

## *KEYWORDS*

*OLAP, Knowledge Discovery from Database, Social Security Fund, Oracle Warehouse Builder, Oracle Discoverer.*

## 1. INTRODUCTION

Decision support systems have a great support to users for making decisions and solving complex problems. They have done this last time, a great success in the professional world. Knowledge Discovery in Database (KDD) is a nontrivial process of analyzing data to extract patterns or knowledge valid, novel and useful to decision-making [3]. KDD is motivated by the massive use of computer systems. The huge volumes of data stored in these systems are improving exponentially, which implies the use of Data Warehousing. A Data Warehouse is defined as a subject-oriented, historical, integrated and non-volatile collection of data, intended for decision support [5]. The Data Warehouse imports data from several operational systems, which are usually heterogeneous. A selection step (the data concerning the subject of the Data Warehouse) and pre-processing of data through ETL tools [7] (Extract, Transform and Load) is needed, this is the integration step. The exploitation and exploration of the data warehouse uses multiple highly sophisticated tools, primarily: OLAP Systems (On-Line Analytical Processing) [7], and Data Mining [3]. We used all the techniques and technologies mentioned above to build a Data Warehouse for the National Social Security Fund –NSSF- in Tunisia.

Social security in Tunisia is entrusted to two agencies: the National Pension and Providence Fund (www.cnrps.nat.tn) and National Social Security Fund –NSSF- (www.cnss.nat.tn). The first deals with the insured public sector and the second deals with the private sector. NSSF manages hundreds of thousands of insured persons and their dependents, according to several schemes and application of laws and regulations. The management of insured uses several computer systems: at the central and regional sites. The complexity and fragmentation of operations on heterogeneous computing systems, physically and logically, make it very difficult and sometimes impossible to extract information and knowledge, and achievement tests for purposes of decision support. To resolve this problem, we designed and implemented a data warehouse that includes all movements of insured accounts. This data warehouse is the core of a system for decision support; it is designed as a multidimensional model based on a factual basis and several dimensions, using the star model. The development of tools for manipulating the warehouse has been done mainly on: the tools of manipulation of multidimensional OLAP structures, tools for extracting reports, charts and analytical formulation of ad-hoc queries (unplanned) adapted to the needs of decision makers.

This article is organized as follows. Section 2 is devoted to the specification of the project and its objectives. Section 3 concerns the design and implementation of data warehouse. Section 4 presents the results of our project. Conclusion and perspectives of this work are discussed in section 5.

## 2. PROJECT SPECIFICATION

An insured is defined as an individual affiliated with a social security scheme. It pays its dues and subsidies and has several benefits. The movements of accounting insured designate its related financial transactions. The NSSF manages a large number of insured persons, exceeding three million. The management of insured persons and their movements accounting, is distributed over several heterogeneous systems: mainframe, n-tier architecture and relational databases. This complicates the analysis for purposes of decision support and especially when confronting the problem of data integrity and consolidation.

Taking into account this problem and before starting the construction of the warehouse, the following targets were set, in collaboration with users and business experts:
- Improved monitoring of monetary transactions covered by social insurance.
- Better knowledge of movement provided by several dimensions: Plan, Nature of Benefits, Time, Location.
- Optimal management of expenditure and revenue.
- Homogenization of heterogeneous data sources: Oracle Database, Excel files, flat files, etc.
- The dynamic analysis of the causes of good functioning and dysfunctions based on different dimensions (the Plan, the Regional Office, Nature of service, period) and crosses between dimensions.
- The dynamic search of correlations between different independent criteria using analytical tools.
- Planned and instantaneous diffusion of information necessary to all decision makers: Scoreboards and Scoring.
- The generation of more detailed summary statements of these movements.

## 3. BUILDING THE DATA WAREHOUSE

At this section, the various stages of design and construction of data warehouse are presented. The study of information system of the NSSF has allowed us to locate the sources of data.

### 3.1 Data Sources.

The following data sources were identified:
- The CICS system: the mainframe, IBM, which manages the financial operations of the NSSF. This system is distributed over the central site and all regional and local subsidiaries. Almost forty COBOL files were selected for data extraction. Each file has a size of approximately 100MB. Earlier versions of files have been used to recover data from previous years. It should be noted that the administration and maintenance of CICS system and its files is a very heavy and complex task, involving several old technologies.

- The Insured Persons Management System: based on a client / server architecture, using Oracle 10g DBMS [8]. The database has almost 6 million rows or records with a size of 120 GB, implemented on a powerful server, SUN. The relational structure of the database has greatly aided and facilitated the task of extracting and manipulating data.

- Several Oracle databases and access files that contain other data necessary for data warehouse.

- Excel files from local and regional structures. These files have typically large sizes of up to 50 MB. These files contain data that are not processed through computer applications (CICS System and Oracle DBMS). The content of these files is extremely important within the NSSF. A set of about 100 Excel files has been processed.

### 3.2. The Design Model of the Data Warehouse

After locating the sources of data, we proceeded to design the Data Warehouse; the star model [11] was adopted, based on a fact table and multiple dimensions. This model is presented in Figure 1 below.

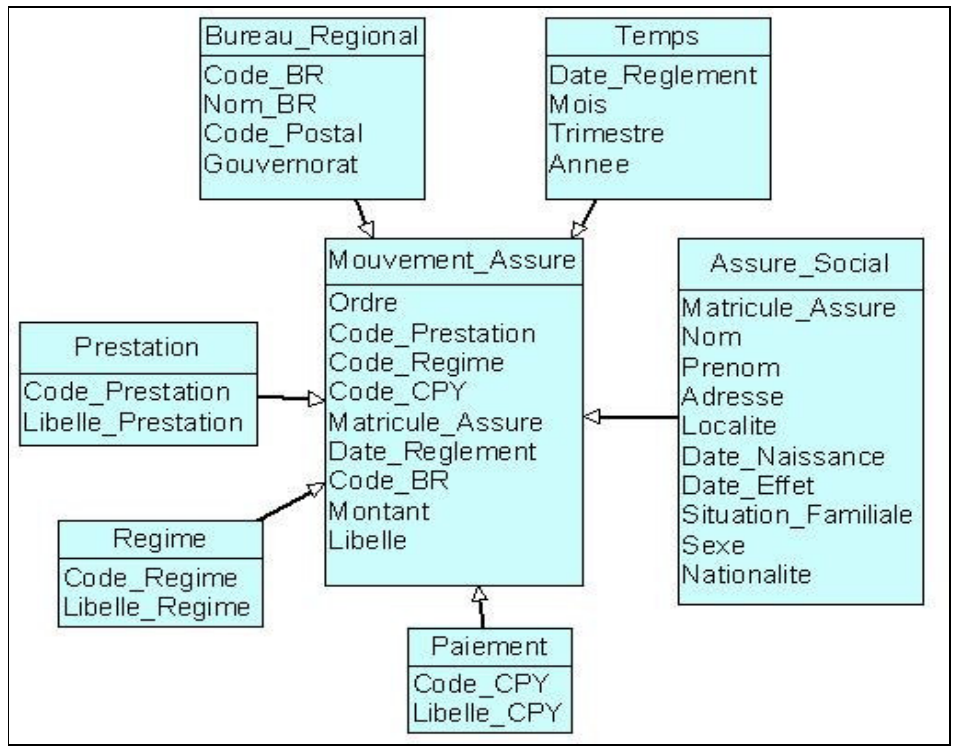

**Figure 1. Data Warehouse Design: The star schema**

The model shown in Figure 2 is the result of the study of the decision objectives, and the existing operational systems that are used to load the warehouse. The different movements of insured accounts are the basic facts; the above dimensions represent different aspects of an element of the basic fact table:

- Social Insured Dimension: This dimension defines the data necessary and available for an insured. It allows seeing the movements of insured persons in relation to their identities and marital status.
- Regional Office Dimension: This dimension can analyze the movements performed by the regional office or by governorate. It uses two hierarchical levels: Regional and Governorate.
- Time Dimension: This dimension helps to visualize the movement provided by the time axis is divided into four significant levels in the business logic of the NSSF: Day, Month, Quarter and Year.
- Payment Dimension: This dimension can classify movements by mode of payment.
- Social Scheme (Regime) Dimension: This dimension brings the types of social schemes offered by the NSSF.
- Benefit (Prestation) Dimension: This dimension presents the different types of benefits of the NSSF.

This multidimensional approach is used to achieve effective decision analysis. To implement the data warehouse, we used Oracle Warehouse Builder 10g [10].

### 3.3 The ETL Process

An ETL (Extract, Transform and Load) has been implemented for loading the data warehouse from various heterogeneous data sources. Oracle Warehouse Builder [10] was used to process data from Oracle databases and Access databases. Regarding the accounting transactions of the central system, we developed an application in Visual Basic, for extracting, transforming and loading data to a relational database initially and then to the multidimensional structures, a sample of this code source, designed to load the data of the regional office dimension is presented below:

```
Private Sub Charger_Bureau_Regional()
      Dim MaComm As ADODB.Command
      Dim cnx As New ADODB.Connection
      Set cnx = New ADODB.Connection
      'Définition de la chaîne de connexion
      cnx.ConnectionString = "Provider=MSDASQL.1;Password=*******;Persist Security
      Info=True;User ID=assurance;Data Source=assurancedns;Mode=ReadWrite"
      'MSDASQL
      'Ouverture de la base de données
      cnx.Open
      If cnx.State = adStateOpen Then
         MsgBox "connection etablie"
      End If
      Set MaComm = New ADODB.Command
      Set MaComm.ActiveConnection = cnx
      Set DataGrid1.DataSource = Adodc3
      DataGrid1.Refresh
      NBENRG.Caption = Adodc3.Recordset.RecordCount
```

```
       dater1 = 0
       matr1 = 0
       CCR1 = 0
        For j = 1 To Adodc3.Recordset.RecordCount + 1
        wcode_br = Adodc3.Recordset.Fields("code_br")
        wnom = Adodc3.Recordset.Fields("nom_br")
        wcode_p = Adodc3.Recordset.Fields("code_postal")
        wgov = Adodc3.Recordset.Fields("Gouvernorat")
        MaComm.CommandText = " insert into bureau_regional VALUES ('" &
    wcode_br & "','" & wnom & "','" & wcode_p & "','" & wgov & "');"
        MaComm.Execute
           g = g + 1
         Adodc3.Recordset.MoveNext
          Next j
           NBENRG1.Caption = Adodc3.Recordset.RecordCount
End Sub
```

The transformation represents an important step in preprocessing and adaptation of data sources to the model of the warehouse. We performed the correction of incorrect values and replacement of missing values by using statistical techniques: Bining [12], Medium [9], and Regression [6]. A particular interest has been allocated to the resolution of conflicts and consolidation of heterogeneous data. Decision support and analytical orientations have been given to a number of attributes and their values (Adding attributes, calculation of aggregates, Standardization ...).

We returned to previous backups of production systems to retrieve data of previous years. Thus, Data Warehouse built spread over a long period which provides more powerful analytical tools. Indeed, the data warehouse built is very large, with a size of 200 GB.

### 3.4. The mechanisms of access and retrieval of knowledge

After building the data warehouse, we created a set of mechanisms and tools for access and exploration of knowledge. Creating an OLAP cube [1, 2, 4] (On Line Analytical Processing) is the first step. Recall that the OLAP cube is a multidimensional logical structure for exploration and navigation in data with OLAP tools and enforcement of various analytical operations with very high information value [7]: ROLL UP, ROTATE, SLICE. The creation of the cube induces the creation of several intermediate calculations supposed to improve the response time thereafter. Figure 2 shows the topology of the cube that we created as generated by Oracle Warehouse Builder.

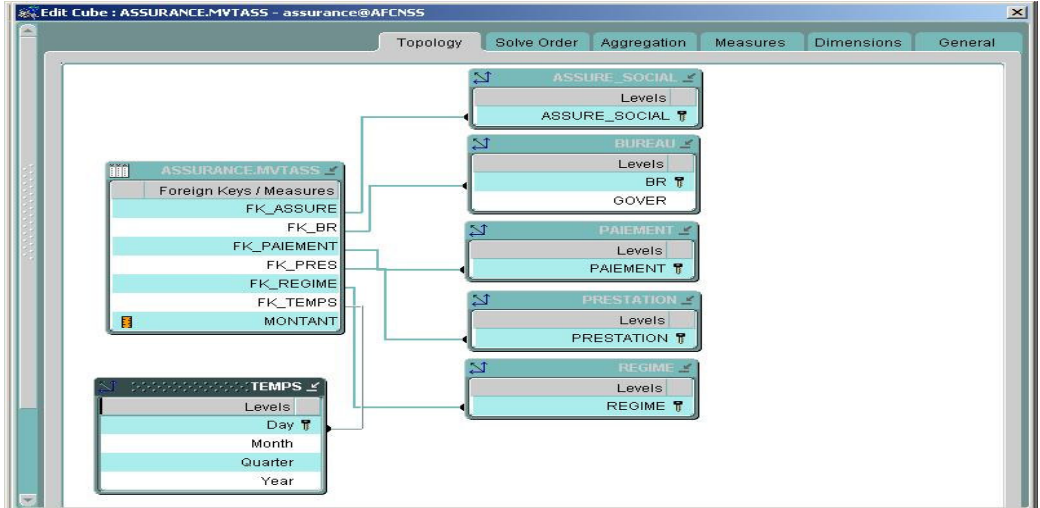

**Figure 2. Cube Topology**

The huge volume of data and the analytical queries affects negatively system performance in terms of response time. To overcome this problem, several materialized views [1] have been created based on the needs and on the tests performed with users and business experts. The principle is the pre-calculation of query results may be asked frequently, these results are stored in dedicated structures to allow quick access. We programmed the reconstruction of those views to each update of the data warehouse. For example, we expose the materialized view **MvtRegPresBr** that stores the calculated amounts of balances of monetary transactions by service, by regional office (We have 41 regional and local offices) and the Scheme dimension. The creation of materialized views implies the achievement of all intermediate computings i.e., including combinations of all variables instantiations, may be performed into users' queries thereafter. This optimizes the response time of the system significantly. This view has the following structure:

| Name | Datatype | Size | Scale | Nulls? | Default Value |
|---|---|---|---|---|---|
| LIBELLE_REGIME | VARCHAR2 | 250 | | No | |
| LIBELLE_PRESTATION | VARCHAR2 | 50 | | No | |
| NOM_BR | VARCHAR2 | 40 | | Yes | |
| SUM(MVTASS.MONTANT) | NUMBER | | | Yes | |

The following source code is the SQL script executed to create the **MvtRegPresBr** view:

```
CREATE MATERIALIZED VIEW "ASSURANCE"."MTPRESTREGBR"
STORAGE ( INITIAL 2M NEXT 2K MAXEXTENTS UNLIMITED)
TABLESPACE "DWCNSS"
BUILD IMMEDIATE
USING INDEX
TABLESPACE "DWCNSS" STORAGE ( INITIAL 2M NEXT 2K MAXEXTENTS
UNLIMITED)
REFRESH FORCE
ON DEMAND
ENABLE QUERY REWRITE
AS
```

```
SELECT
REGIME.LIBELLE_REGIME,PRESTATION.LIBELLE_PRESTATION,BUREAU_REGIONAL.NOM_BR,SU
M(MVTASS.MONTANT)
FROM BUREAU_REGIONAL,PRESTATION,MVTASS,REGIME
WHERE
(MVTASS.CODE_BR=BUREAU_REGIONAL.CODE_BR) AND
(MVTASS.CODE_PRESTATION=PRESTATION.CODE_PRESTATION) AND
(MVTASS.CODE_REGIME=REGIME.CODE_REGIME)
GROUP BY LIBELLE_REGIME,BUREAU_REGIONAL.NOM_BR,LIBELLE_PRESTATION;
```

On another level, we have used Oracle Discoverer [10], a tool of Oracle Business Intelligence suite, which exists in two versions: Administrator and Desktop, to create reports of exploration data warehouse. These reports can be manipulated interactively by the user and allow him to express his choice dynamically, which is the main objective of developing OLAP solutions. In addition to that, users can create their own reports or charts and navigate into the hierarchies of dimensions. The interface between Discoverer and the Oracle Data warehouse is done through the layers EUL (End User Layers) created with the Administrator tool. In Section 4, we present examples of results generated.

## 4. RESULT

We expose the results and cases of use of the Data Warehouse, which combined different date sources for the first time in their life cycles.

- **Figure 3** below shows a sample of the report: Distribution of social benefits balance amounts by Governorate (we have 24 governorates) and type of Benefits (we have 8 Social Security Benefits) for a user defined time interval.

- **Figure 4** shows a sample of the report of the amounts of benefits according to several dimensions and axes of analysis: Time, Regime (We have 6 regimes), Governorate, type of Benefits. We note the high level of usability of the report which allows the user to modify all parameters depending on his information needs. That was our goal from the beginning. Before the construction of OLAP, it was very difficult and sometimes impossible to extract the information contained in the report of Figure 4, necessary for control and monitoring. According to a study that we conducted, it required 3 or 4 days of treatment with the intervention of at least one computer scientist. With OLAP, the repot takes only 60 minutes of execution, with the possibility of intervention by the user to change input parameters and see results in real time. The following source code is the SQL script executed to extract this report on relational data bases without creating OLAP Cubes:

    ```
    SELECT REGIME.LIBELLE_REGIME, PRESTATION.LIBELLE_PRESTATION,
    BUREAU_REGIONAL.NOM_BR, SUM(MVTASS.MONTANT)
    FROM BUREAU_REGIONAL, PRESTATION, MVTASS,REGIME
    WHERE (MVTASS.CODE_BR = BUREAU_REGIONAL.CODE_BR)
    AND (MVTASS.CODE_PRESTATION = PRESTATION.CODE_PRESTATION)
    AND (MVTASS.CODE_REGIME = REGIME.CODE_REGIME)
    GROUP BY LIBELLE_REGIME, BUREAU_REGIONAL.NOM_BR,  LIBELLE_PRESTATION;
    ```

- A fundamental characteristic of OLAP systems is that they can see the data at all levels of abstraction: From highest to lowest. Figure 5 provides an extract of the allocation of accounting transactions of a covered person (the most basic level of abstraction). With the possibility of changing parameters (Regime, insured person, level of abstraction ...) interactively as shown in Figure 5. We've hidden the identity of the insured person for reasons of protection of private data.

**Figure 3. Distribution of social benefits balance amounts by governorate, by regional office and benefit type**

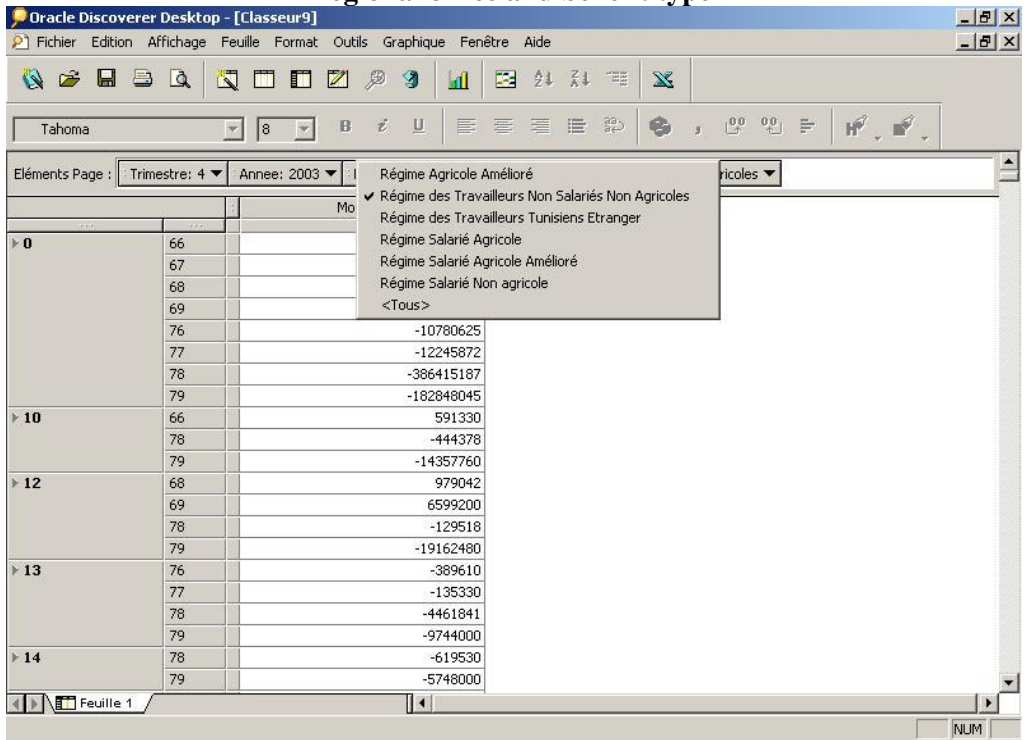

**Figure 4. Distribution of social benefits balance amounts across multiple dimensions with real time update capabilities**

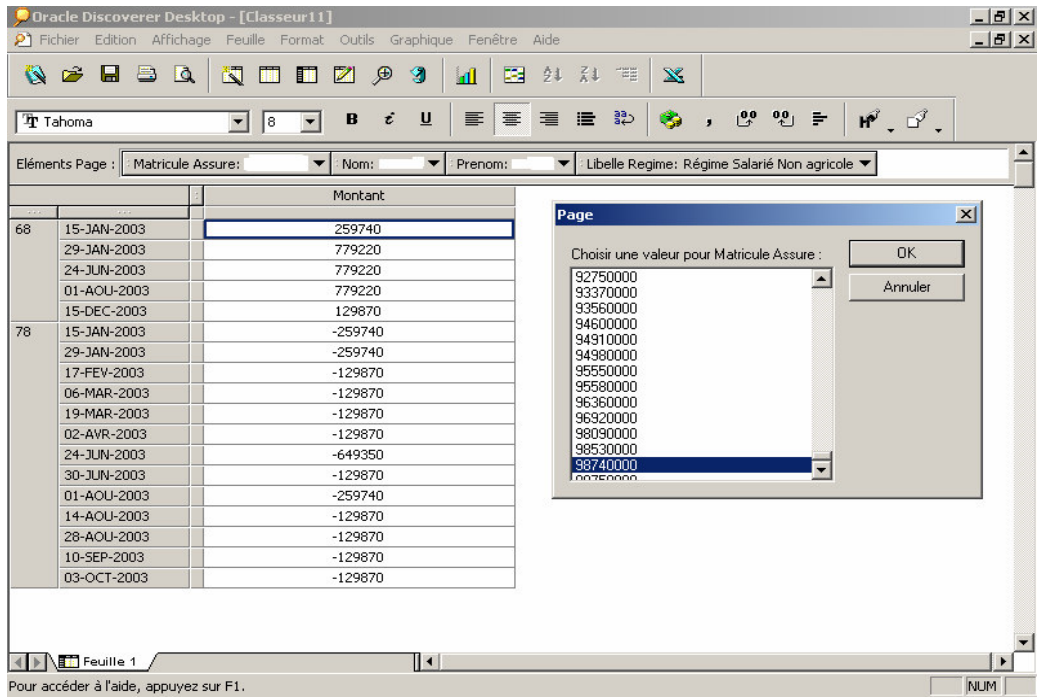

**Figure5. Social benefits amounts of a well defined social insured person
(Lowest level of abstraction)**

- The display of information is a key factor for the user in interpreting the results. Our system allows the retrieval and the analysis of results in many formats. Figure 6 shows a diagram presenting the balance amounts by governorate.

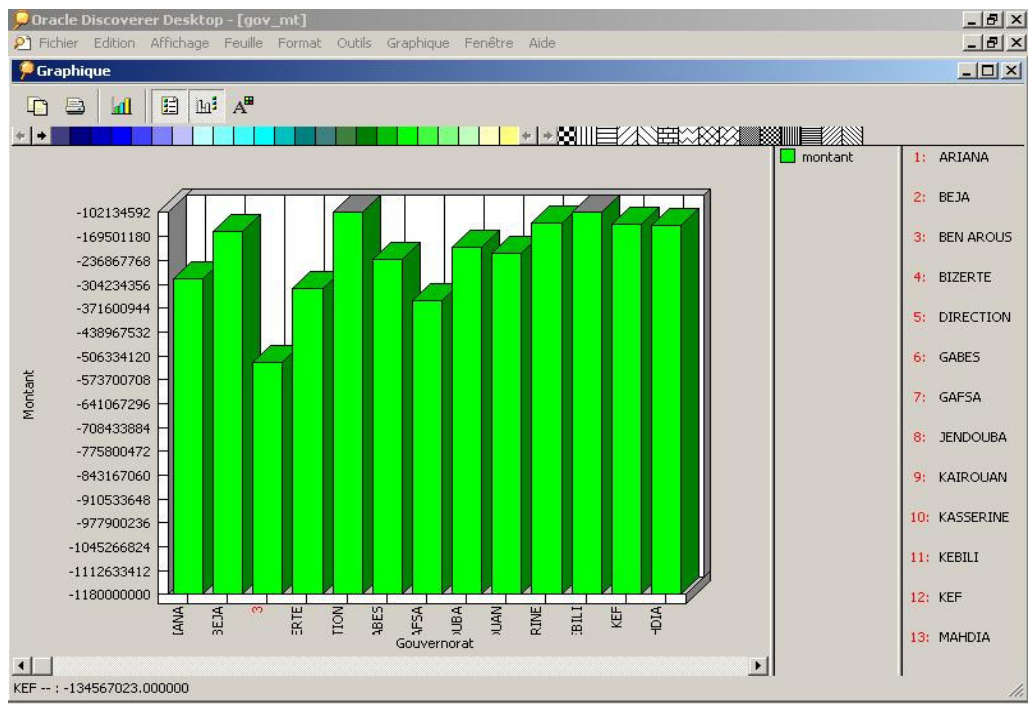

**Figure 6. Diagram of social benefits balance amounts by governorate**

- At figure 7, a drill down operation has been applied on the Benefits Dimension which gave the allocation of accounting balances by type of benefit, and we show that the system can move from one level to other easily. All these operations can be performed dynamically and interactively by the user. The computational time consumed to extract these diagrams is of the order of 100 minutes or more, it depends deeply on the analysis period designated by the user. It should be noted that these diagrams could not be generated before the construction of the OLAP system: lack of integration of data and advanced analytics. Moreover, the analytical decision support query execution is a high computational load which may affect the performance of production systems, given the absence of computing units dedicated to decision support.

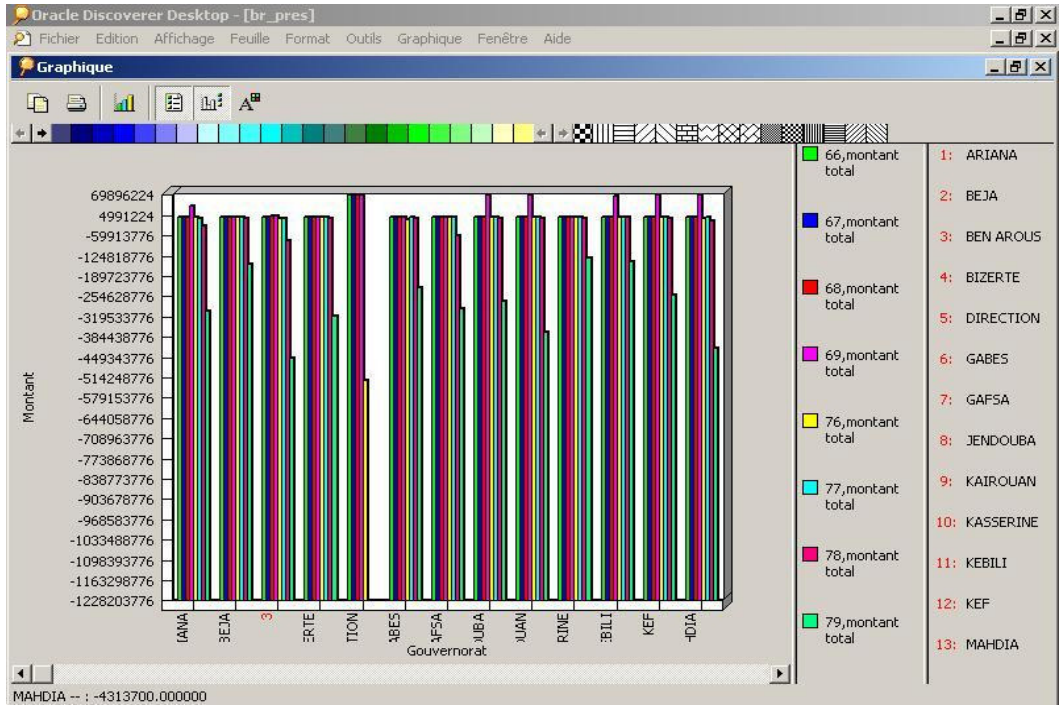

**Figure 7. Drill Down on Benefit Dimension (Figure 6)**

- The last example shown in figures 8 and 9, shows the techniques of handling the time dimension and the possibility of extracting information by day, month, quarter and year (DRILL UP and DRILL DOWN) under dynamic choices of the user. We note that the system allows you to view multiple levels of abstraction of the time dimension on the same interface.

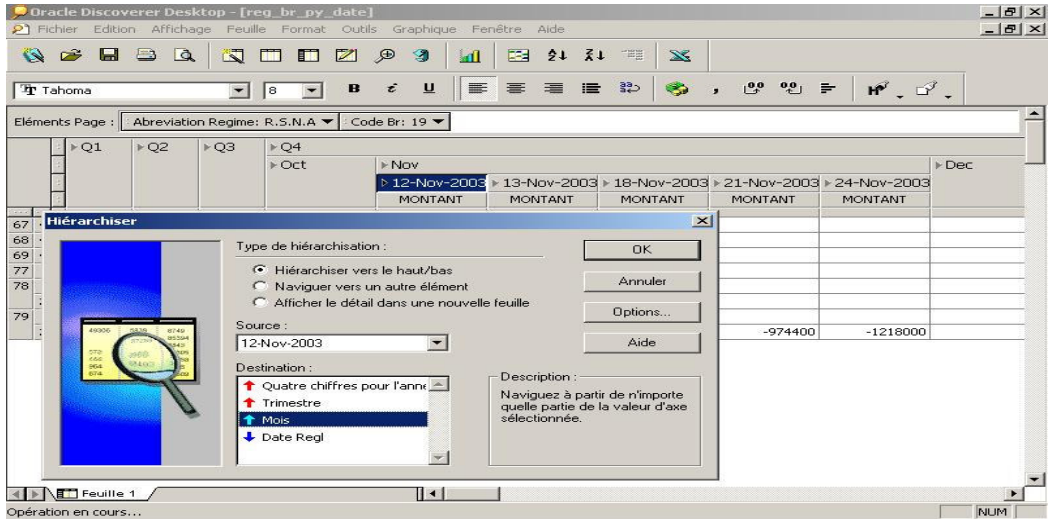

**Figure 8. Handling the Time Dimension for real time update of information**

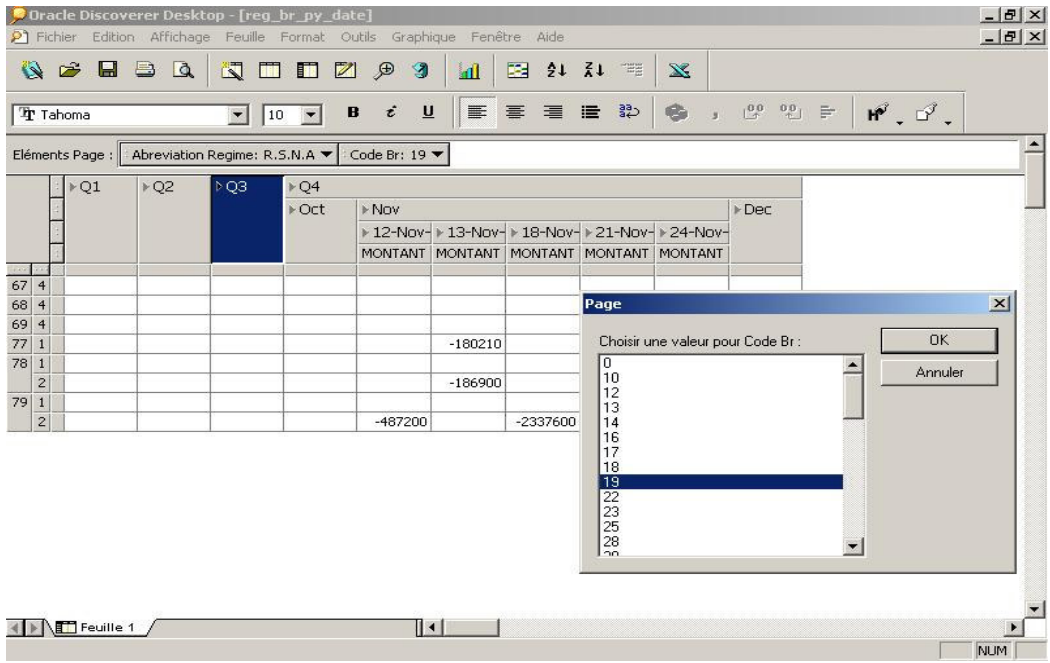

**Figure 9. Changing other parameters of Figure 8.**

At this section, we exposed a sample of results that can be generated by the data warehouse tools that we built. This includes all accounting transactions of insured persons.

## 5. CONCLUSION

In this paper, we addressed the problem of creating a solution for decision support within the National Social Security Fund of the Republic of Tunisia. To achieve this goal, we designed and built a Data Warehouse of accounting transactions of insured persons, whose data from production systems. The process of Knowledge Discovery from Data has been applied in all its steps, including pre-processing, transformation and consolidation. This step was very difficult given the heterogeneity of data sources (Cobol files, relational databases, Access ...) and the multitude of problems we encountered during the consolidation of data (missing data, wrong data ...), it represented approximately 50% of the total period of the project. We used the Oracle Warehouse Builder ETL tool (Extract, Transform and Load). We also implemented our own programs to deal with some complex data sources. A multidimensional approach has been adopted for the creation of OLAP cubes and materialized views according to the preliminary studies and the initial objectives. Several tools have also been made available to users for the exploration of the Data Warehouse and retrieval of relevant knowledge useful in making strategic decisions in different formats in a dynamic and interactive way. Users and experts of the NSSF, always hampered by limited decision support resources, expressed a high level of satisfaction and commitment to use the Data Warehouse. We have mentioned some samples of the results extracted in Section 4 of this Article.

The prospects of this work remain the creation of solutions for other fields the NSSF and the implementation of Data Mining techniques for data analysis and knowledge extraction.

**Authors**

Mohamed Salah **GOUIDER** is associate professor at the University of Tunis – Tunisia. He now has thirty years experience in the field of databases and in recent years in the field of Data Warehouse and Data Mining. He earned his doctorate at the Faculty of Science - University of Nice - France in 1983. He is currently a consultant top management in several public and private enterprises. He has extensive experience in several countries: France, Tunisia, Kuwait, Qatar and Benin. His recent research is focused mainly in the extraction of knowledge from data in the medical and financial, as well as the optimization algorithms of the Data Warehouse and Data Mining.

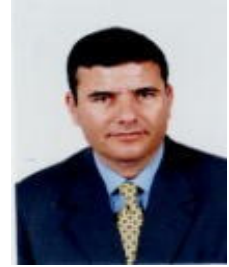

Amine **FARHAT**, Master of Science in Computer Science, is a researcher and Ph.D. Student at the BESTMOD Laboratory, University of Tunis. He is assistant professor in Computer Science in the ISG of Tunis. He is also a senior computer science analyst Engineer, Head of Software Development and Decision Support Systems Projects, in a public Office. His research Interests are mainly Knowledge Discovery in Databases, Data mining algorithms, Artificial Intelligence and Data warehousing.

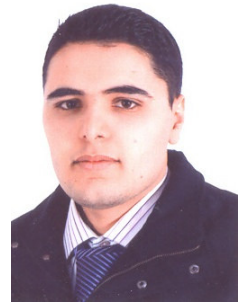